\documentclass[letterpaper, preprint, paper,11pt]{AAS}	

\usepackage{bm}
\usepackage{amsmath}
\usepackage{subfigure}
\usepackage[colorlinks=true, pdfstartview=FitV, linkcolor=black, citecolor= black, urlcolor= black]{hyperref}
\usepackage{overcite}
\usepackage{footnpag}			      	
\usepackage{upgreek}
\usepackage{makecell}

\PaperNumber{22-556}

\begin{document}

\title{LISA POINT-AHEAD ANGLE CONTROL FOR OPTIMAL TILT-TO-LENGTH NOISE ESTIMATION}

\author{Niklas Houba\thanks{Doctoral Candidate, Airbus Space Systems, Airbus Defence and Space GmbH, 88090 Immenstaad am Bodensee, Germany; registered as a Dr.-Ing. candidate at the University of Stuttgart.}, 
Simon Delchambre\thanks{DFACS Systems Engineer, Airbus Space Systems, Airbus Defence and Space GmbH, 88090 Immenstaad am Bodensee, Germany.}, 
Tobias Ziegler\thanks{Lead Systems Engineer, Airbus Space Systems, Airbus Defence and Space GmbH, 88090 Immenstaad am Bodensee, Germany}, 
Gerald Hechenblaikner\thanks{Instrument System Engineer, Airbus Space Systems, Airbus Defence and Space GmbH, 88090 Immenstaad am Bodensee, Germany.}, 
and Walter Fichter\thanks{Director of the
Institute of Flight Mechanics and Control, Institute of Flight Mechanics and Control, University of Stuttgart, 70569 Stuttgart, Germany.}}


\maketitle{}

\begin{abstract}
The Laser Interferometer Space Antenna (LISA) mission features a three-spacecraft long-arm constellation intended to detect gravitational wave sources in the low-frequency band up to 1 Hz via laser interferometry. The paper presents an open-loop control strategy for point-ahead angle (PAA) correction required to maintain the optical links of the moving constellation. The control strategy maximizes periods between adjustments at the constellation level and is shown to be optimal from the perspective of estimating and correcting tilt-to-length (TTL) coupling. TTL is a noise source that couples angular spacecraft jitter and jitter of optical subassemblies with longitudinal interferometer measurements. Without precise TTL noise estimation and correction, TTL coupling fundamentally limits the detector's sensitivity.
\end{abstract}

\section{Introduction}
High precision pointing missions, such as LISA, require a sophisticated pointing control design to achieve their scientific objectives. In LISA, the angular jitter of the spacecraft (SC) cannot be completely averted by the feedback system, contributing to measurement noise in the science data via TTL coupling. Recent analyses show that TTL coupling fundamentally limits the detector's sensitivity. Consequently, estimation and correction of the residual jitter noise contribution are required and expected to be performed during offline data processing. To meet the estimation accuracy requirement, the estimation window of TTL estimation campaigns must be chosen sufficiently large. The introduction explains how PAA correction affects TTL estimation campaigns and why optimal control of the PAA mechanism is essential for TTL noise reduction in LISA.
Section \ref{sec:2} derives analytical expressions to describe temporal PAA variations for Keplerian LISA orbits. Section \ref{sec:3} presents an open-loop control strategy for PAA readjustment. The paper concludes with a robustness analysis of the control concept regarding variations of the initial
ecliptic longitude and the initial orientation of the constellation.
\subsection{LISA Mission}
\begin{figure}[b!]
	\centering
  \includegraphics[trim=65 10 10 28, clip,width=0.62\textwidth]{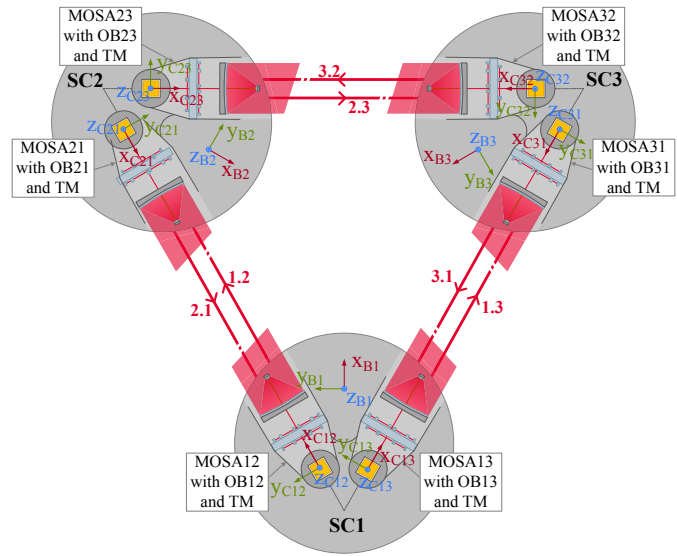}
	\caption{Illustration of the LISA Constellation }
	\label{im:LISAConstellation}
\end{figure}

LISA is the proposed concept of a space mission to establish the new field of gravitational wave astronomy, observing highly energetic events within the entire event horizon of the universe. The triangular constellation of three spacecraft moves in an Earth-trailing orbit around the Sun. The nominal distance between the remote spacecraft is about 2.5 million kilometers. A schematic illustration of the constellation is given in Figure \ref{im:LISAConstellation}.  It can be seen that each spacecraft carries two Moving Optical Subassemblies (MOSAs), where each MOSA comprises an optical bench (OB), a telescope, and a free-falling test mass (TM) \cite{obdev}.  LISA  aims at resolving gravitational wave-related distance changes between remote test masses \cite{LMWP}. To obtain the test mass to test mass distance, various interferometric measurements are performed onboard. These are combined in post-processing as synthetic Michelson interferometer signals using the Time-Delay Interferometry (TDI) concept. TDI is required to suppress laser frequency fluctuations present in the optical bench measurements, which is a general problem of spaceborne unequal long arm interferometers, such as LISA \cite{TDITINTO2005}.  Unfortunately, TTL noise is not suppressed by TDI. It disturbs the TDI output and needs to be handled by a dedicated noise reduction algorithm (see References~\citenum{houba} and \citenum{houba2}).
\subsection{Tilt-to-Length Coupling} 
\begin{figure}[]
	\centering
 \includegraphics[trim=300 180 60 210, clip,width=0.5\textwidth]{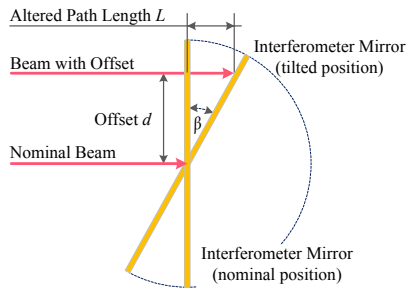}
 	\caption{TTL-Induced Path Length Change}
	\label{im:ttlsimple}
\end{figure}

Angular spacecraft jitter and jitter of the spacecrafts' optomechanical subassemblies induce noise in LISA's longitudinal interferometer measurements via TTL coupling. Figure \ref{im:ttlsimple} provides a descriptive example of geometric TTL coupling: An interferometer mirror tilt $\beta$ results in a TTL-induced change in path length according to the offset $d$ of a laser beam relative to its nominal position. Assuming small angles and linearity, $ L=d\cdot\beta$ gives the altered path length (one-way) due to TTL coupling. 

TTL can be modeled by the product of a coupling factor, termed $C$, and the jitter of the spacecraft. The coupling factor quantifies the tilt's impact on the altered path length. In the example of Figure \ref{im:ttlsimple}  the coupling factor is given by $C=\frac{\partial L}{\partial \beta}=d$\enlargethispage{\baselineskip}. In LISA, the TTL coupling sources are manifold. For instance, misalignment of optical components can introduce undesirable laser beam offsets. An example of non-geometric TTL is laser wavefront distortion. The wavefront of an emitted laser beam is not spherical in reality but exhibits deformations that are preserved over the propagation distance to the receiving spacecraft. If the emitting spacecraft jitters, the deformed wavefront at the receiving spacecraft detector leads to variations in the measured phase, thus causing TTL noise in the measurement. Geometric and non-geometric TTL sources are analyzed in References~\citenum{hartig2022geometric} and \citenum{hartig2021nongeometric}. They are summarized by an effective TTL coupling factor. Knowing this coupling factor as well as the angular jitter allows removing TTL from the TDI output. Since no direct measurements of the TTL coupling factors are available, the task is to estimate them based on the TDI output.

The ideal TTL correction requires the estimation error to be zero. Otherwise, TTL noise remains in the TDI signals after TTL subtraction. To correct TTL sufficiently, estimation accuracies for TTL coupling factors must be met, which are presented in Reference~\citenum{houba2}, for example. Assuming time-invariant coupling factors, the estimation accuracy increases with an increasing number of measurement samples, and thus extended estimation windows. The impact of PAA adjustment on TTL estimation windows will be explained subsequently.
\subsection{Point-Ahead Angle}
In LISA, each optical bench must direct its outgoing laser beam towards the position where the receiving optical bench will be after the beam has crossed the inter-spacecraft distance of 2.5 million km during a time of approximately 8.3 sec. This ensures that the remote spacecraft can detect the laser beam. The receiving optical bench thus observes a laser beam from the direction determined by the past position of the transmitting spacecraft when the laser beam was sent. The different orientation between an optical bench's incoming and outgoing laser beam is denoted as PAA. It considers relative position changes of the receiving spacecraft during laser propagation. 

Orbital dynamics causes variations in the PAA to be compensated by the one-axial PAA mechanism (PAAM), which adjusts the orientation of the outgoing laser beams on each optical bench as described in Reference~\citenum{paamdesign1}. The PAAM is a rotatable mirror placed in the path of the outgoing laser beam. Unfortunately, discrete PAAM actuations are expected to cause TTL coupling factor changes. The reasons for this are:  
\vspace{5pt}
\begin{enumerate}
\item \textbf{Change of field angle}: After a discrete PAAM actuation, the laser beam passes through a different field region of the telescope optics, which leads to a change in the TTL coupling factors through the dependence of the coupling factors on the field region. If the PAAM is continuously actuated, the total change of the field angle could be smaller, avoiding considerable jumps of the beam in the field. Nevertheless, the paper assumes discrete step-and-stare operation of the PAAM, which corresponds to the current system design baseline.

\item  \textbf{Rotation axis of the PAAM}: 
Suppose that the center of rotation of the PAAM mirror surface is not precisely located in the mirror plane and the beam or the beam incident on the PAAM has a lateral offset from the center of rotation. In that case, there will be a change in the TTL coupling factors when the PAAM is rotated. The amount of change depends on the accuracy of the (lateral) alignment of the mechanism concerning the beam path on the optical bench and the accuracy of the mechanism, \emph{i.e.}, the accuracy of the positioning of the rotation axis on the surface of the mirror.
\item  \textbf{Pupil position of the beam}: If the beam's pupil is not precisely placed on the surface of the PAAM, there is a jump of the TTL coupling factor to be expected after PAAM actuation. By definition of the TTL effect, coupling factors change when the beam is rotated, but not rotated concerning the pupil plane. This effect should be small compared to the effects mentioned because the current design places the PAAM exactly on the pupil plane.
\end{enumerate}
\vspace{5pt}
Changes in TTL factors during TTL estimation campaigns should be avoided to decrease estimation errors and increase the TTL subtraction performance considering that no dynamical model is available to predict the evolution of TTL factors. As presented in Reference~\citenum{houba}, the dominant TTL effect in the long-arm interferometers of the LISA constellation can be described by 24 different TTL coupling factors. These have to be simultaneously estimated when the estimation is performed based on residual spacecraft jitter without dedicated maneuvers. Therefore, coordinating the PAAM actuation between the six optical benches is reasonable to avoid changes of individual coupling factors during the estimation. Before presenting the actuation strategy, analytical expressions are derived that quantify the temporal variations of the PAA.

\section{Mathematical Analysis of Temporal Point-Ahead Angle Variations}\label{sec:2}
\vspace{5pt}
Each optical bench of the constellation transmits a laser beam to the remote spacecraft. Consequently, six PAA must be taken into account for LISA. This section contains the derivation of the analytical PAA expressions and the numerical simulation of temporal PAA variations for Keplarian orbits.

\subsection{Derivation of In-Plane and Out-of-Plane Point-Ahead Angles}
\begin{figure}[]
	\centering
 \includegraphics[trim=150 125 0 370, clip,width=0.55\textwidth]{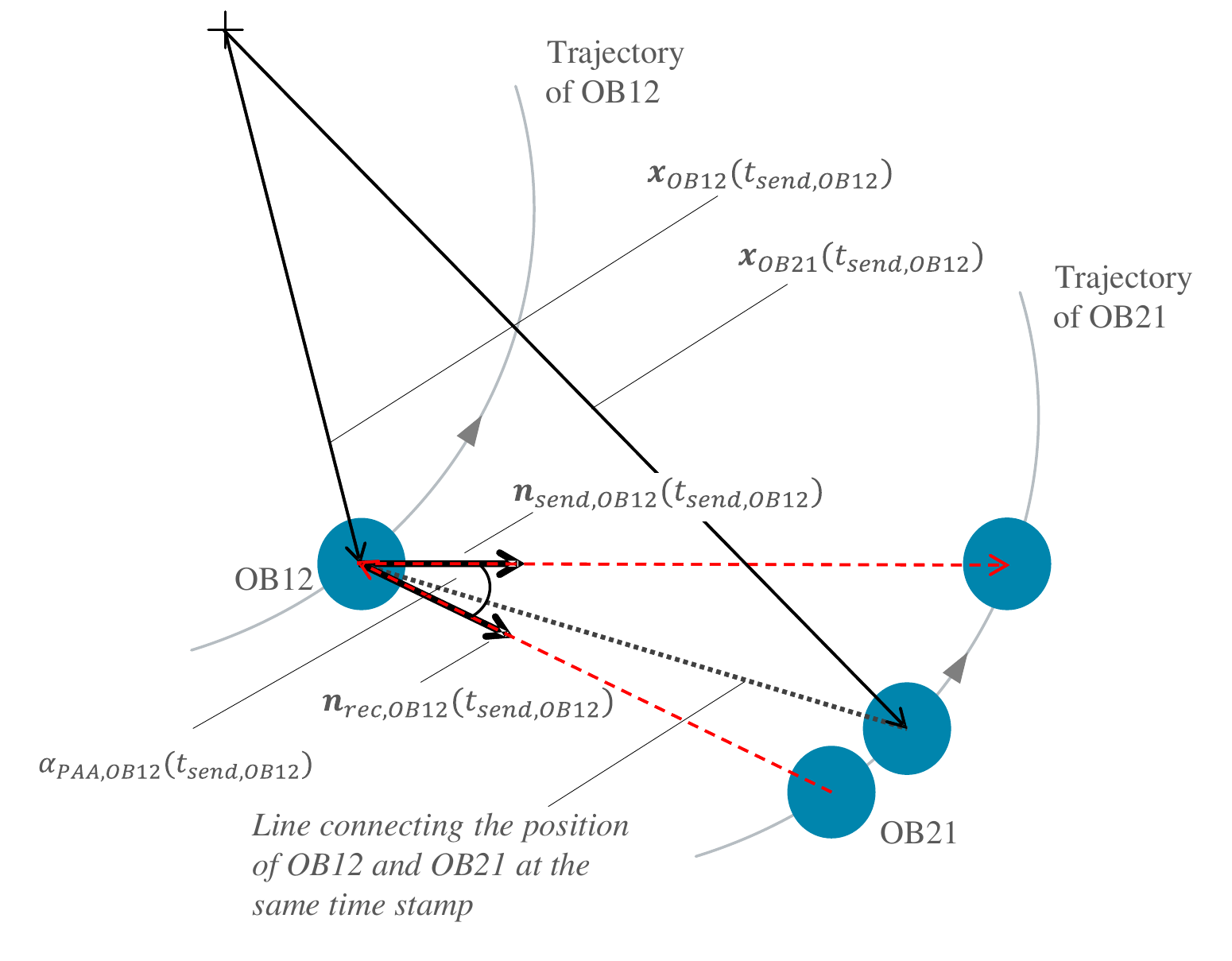}
 	\caption{PAA of OB12}
	\label{im:PAA}
\end{figure}

\begin{figure}[]
	\centering
 \includegraphics[trim=150 150 10 360, clip,width=0.55\textwidth]{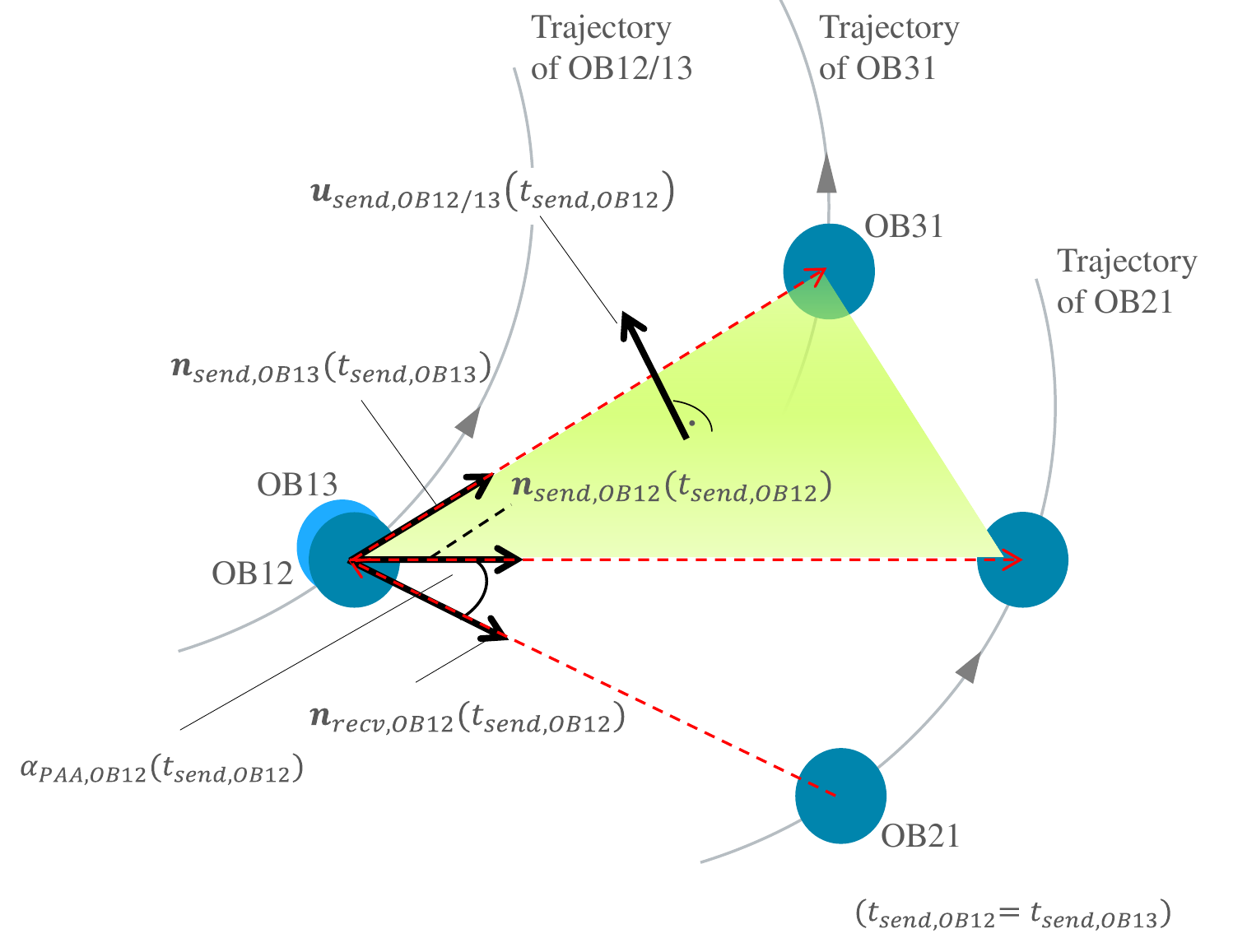}
 	\caption{Plane Definition for OB12}
	\label{im:PAAPlanes}
\end{figure}

The PAA is the angle between an optical bench's incoming and outgoing laser beam. Figure \ref{im:PAA} illustrates the PAA definition using OB12 as an example. OB12 is assumed to be a point mass moving along the associated trajectory defined by the orbit of SC1. The depicted position of OB12 corresponds to that at time $t=t_{send,OB12}$, which is when OB12 transmits its laser beam to the distant optical bench, \emph{i.e.}, OB21. The outgoing laser beam is shown by the red dashed arrow starting at OB12 and directed to the position of OB21 at the time of reception $t_{rec,OB21} > t_{send,OB12}$. If OB12 sent the laser beam to the position of OB21 at $t_{send,OB12}$ (black dotted line in Figure \ref{im:PAA}), OB21 would not detect the laser beam because it has already moved along the trajectory during 2.5 million km of laser propagation.
The incoming laser beam of OB12 is represented by the red dashed arrow directed at OB12 at $t_{send,OB12}$. Its starting point corresponds to the position of OB21 at the time of sending $t_{send,OB21}<t_{send,OB12}$ with analogous reasoning. The angle $\alpha_{PAA,OB12}$ between the laser beams shown in red is the PAA to be considered by OB12 at $t_{send,OB12}$.
Its mathematical definition is given by:
\begin{equation}
\alpha_{P A A, O B 12}\left(t_{{send,OB12 }}\right)=\arccos\left( {\boldsymbol{n}_{{send,OB12 }}\left(t_{{send,OB12 }}\right) \cdot \boldsymbol{n}_{{rec,OB12 }}\left(t_{{send,OB12 }}\right)}\right) \label{eq:PAAGeneral}
\end{equation}
The unit vector $\boldsymbol{n}_{{send,OB12 }}\left(t_{{send,OB12 }}\right)$ describes the orientation of the outgoing laser beam of OB12 at $t_{ {send,OB12 }}$ based on the position vectors $\boldsymbol{x}_{OB12}$ of OB12 and $\boldsymbol{x}_{OB21}$ of OB21 given in the inertial frame:
\begin{equation}
    \boldsymbol{n}_{{send,OB12 }}\left(t_{{send,OB12 }}\right) = \frac{\boldsymbol{x}_{OB21}(t_{ {rec,OB21 }})-\boldsymbol{x}_{OB12}(t_{ {send,OB12 }})}{\left\|\boldsymbol{x}_{OB21}(t_{ {rec,OB21 }})-\boldsymbol{x}_{OB12}(t_{ {send,OB12 }})\right\|_2}\label{eq:Defnsend}
\end{equation}
The unit vector $\boldsymbol{n}_{{rec,OB12 }}\left(t_{{send,OB12 }}\right)$ describes the reverse direction of the incoming laser beam received by OB12 at $t_{ {send,OB12 }}$:
\begin{equation}
    \boldsymbol{n}_{{rec,OB12 }}\left(t_{{send,OB12 }}\right) = \frac{\boldsymbol{x}_{OB21}(t_{ {send,OB21 }})-\boldsymbol{x}_{OB12}(t_{ {send,OB12 }})}{\left\|\boldsymbol{x}_{OB21}(t_{ {send,OB21 }})-\boldsymbol{x}_{OB12}(t_{ {send,OB12 }})\right\|_2}\label{eq:Defnrec}
\end{equation}
It is common practice to split up the PAA $\alpha_{PAA,OB12}$ into an out-of-plane PAA $\eta_{PAA,OB12}$ and an in-plane PAA $\phi_{PAA,OB12}$. As will be seen later, the variations of $\phi_{PAA}$ of any optical bench are much lower than variations of $\eta_{PAA}$, such that only the constellation's out-of-plane PAA need to be considered in the proposed control strategy. Figure \ref{im:PAAPlanes} highlights the plane required to derive the analytical expressions for $\eta_{PAA}$ and $\phi_{PAA}$, again taking OB12 as an example. 
At time $t=t_{send,OB12}$ the plane is described by sending directions $\boldsymbol{n}_{send,OB12}$ of OB12 and $\boldsymbol{n}_{send,OB13}$ OB13, defined as:
\begin{equation}
    \boldsymbol{n}_{{send,OB13 }}\left(t_{send,OB13}=t_{{send,OB12 }}\right) = \frac{\boldsymbol{x}_{OB31}(t_{ {rec,OB31 }})-\boldsymbol{x}_{OB13}(t_{ {send,OB13 }})}{\left\|\boldsymbol{x}_{OB31}(t_{ {rec,OB31 }})-\boldsymbol{x}_{OB13}(t_{ {send,OB13 }})\right\|_2}\label{eq:Defnsend13}
\end{equation}
The out-of-plane PAA $\eta_{PAA,OB12}$ corresponds to the portion of $\alpha_{PAA,OB12}$ that is orthogonal to the plane depicted, see Eq. \eqref{eq:oopPAA}:
\begin{equation}
\eta_{PAA,OB12}(t_{send,OB12}) =\arcsin \boldsymbol{u}_{send,OB12/13}(t_{send,OB12})\cdot\boldsymbol{n}_{rec,OB12}(t_{send,OB12})  \label{eq:oopPAA} 
\end{equation}
Here, $\boldsymbol{u}_{send,OB12/13}(t)$ referts to the plane's normal vector, given by:
\begin{equation}
\boldsymbol{u}_{send,OB12/13}(t)=\frac{\boldsymbol{n}_{send,OB12}(t)\times\boldsymbol{n}_{send,OB13}(t)}{\left\|\boldsymbol{n}_{send,OB12}(t)\times\boldsymbol{n}_{send,OB13}(t)\right\|}
\end{equation}
To obtain the in-plane PAA $\phi_{PAA,OB12}$, the projection of $\boldsymbol{n}_{rec,OB12}(t)$ on the plane $\boldsymbol{u}_{send,OB12/13}(t)$ is required:
\begin{align}
{\boldsymbol{n}}_{rec,OB12,proj}(t) &= \frac{\tilde{\boldsymbol{n}}_{rec,OB12,proj}(t)}{\left\|\tilde{\boldsymbol{n}}_{rec,OB12,proj}(t)\right\|_2},\,\textrm{with} \\
    \tilde{\boldsymbol{n}}_{rec,OB12,proj}(t) &= \boldsymbol{n}_{rec,OB12}(t) - s(t)\cdot\boldsymbol{u}_{send,OB12/13}(t),\,\textrm{and}\\
    s(t) &= \boldsymbol{n}_{rec,OB12}(t)\cdot\boldsymbol{u}_{send,OB12/13}(t)
\end{align}
Then, for $\phi_{PAA,OB12}$, it applies:
\begin{equation}
    \phi_{PAA,OB12}(t_{send,OB12})=\arccos{\left(\boldsymbol{n}_{rec,OB12,proj}(t_{send,OB12})\cdot\boldsymbol{n}_{send,OB12}(t_{send,OB12})\right)} \label{eq:ipPAA}
\end{equation}
In-plane and out-of-plane PAA for the remaining five optical benches are obtained by cyclic permutation of the indices.
\subsection{LISA Orbit}
The orbit derived in Reference~\citenum{2004orbit} is used for the numerical evaluation of LISA's PAA.
Perturbations, \emph{e.g}., from other planets, are not considered so that the three LISA spacecraft follow Keplerian orbits, given by Eqs. \eqref{eq:orbx} to \eqref{eq:orbz} for time $t$ in sec and $i=1,2,3$.
The authors of Reference~\citenum{2004orbit} chose a heliocentric ecliptic coordinate system, where the origin of the coordinate system and the Sun's center of gravity coincide. The x-axis points towards the vernal equinox; the z-axis is parallel to the Earth's angular momentum vector; the y-axis completes the right-hand system. Note that Eqs. \eqref{eq:oopPAA} and \eqref{eq:ipPAA} are independent from the orbit model chosen for the LISA constellation.
\begin{align}
&\begin{aligned}
x_i(t)&= R \cos (\alpha)+\frac{1}{2} e R(\cos (2 \alpha-\beta_i)-3 \cos (\beta_i)) +\frac{1}{8} e^{2} R(3 \cos (3 \alpha-2 \beta_i)\\&-10 \cos (\alpha) -5 \cos (\alpha-2 \beta_i))\label{eq:orbx} \end{aligned}\\
&\begin{aligned}
y_i(t)&= R \sin (\alpha)+\frac{1}{2} e R(\sin (2 \alpha-\beta_i)-3 \sin (\beta_i)) +\frac{1}{8} e^{2} R(3 \sin (3 \alpha-2 \beta_i)\\&-10 \sin (\alpha)+5 \sin (\alpha-2 \beta_i))\label{eq:orby}\end{aligned} \\
&\begin{aligned}
z_i(t)&=-\sqrt{3} e R \cos (\alpha-\beta_i) +\sqrt{3} e^{2} R\left(\cos ^{2}(\alpha-\beta_i)+2 \sin ^{2}(\alpha-\beta_i)\right)\label{eq:orbz}, 
\end{aligned}
\end{align}
with $R=1$ AU: distance between the constellation and the Sun,
\newline $e={L}/({2\sqrt{3}R})$: eccentricity of each LISA spacecraft orbit,
\newline $L = 2.5\cdot 10^9$ m: nominal LISA arm length,
\newline $\alpha = 2\pi f_m t + \kappa$: orbital phase of the constellation's center of mass, which can be related to the eccentric anomaly via Kepler's equation as provided in Reference~\citenum{2004orbit},
\newline $f_m=1$ cycle/year: orbital frequency,
\newline $\kappa$: initial ecliptic longitude of the constellation's center of mass,
\newline $\beta = 2\pi(i-1)/3 + \lambda$: relative phase of a spacecraft within the constellation,
\newline $\lambda$: initial orientation of the constellation defined in Reference ~\citenum{2004orbit}.
\subsection{Numerical Simulation}
\begin{figure}
\centering
\begin{minipage}{.5\textwidth}
  \centering
  \includegraphics[trim=90 260 100 180, clip,width=1\textwidth]{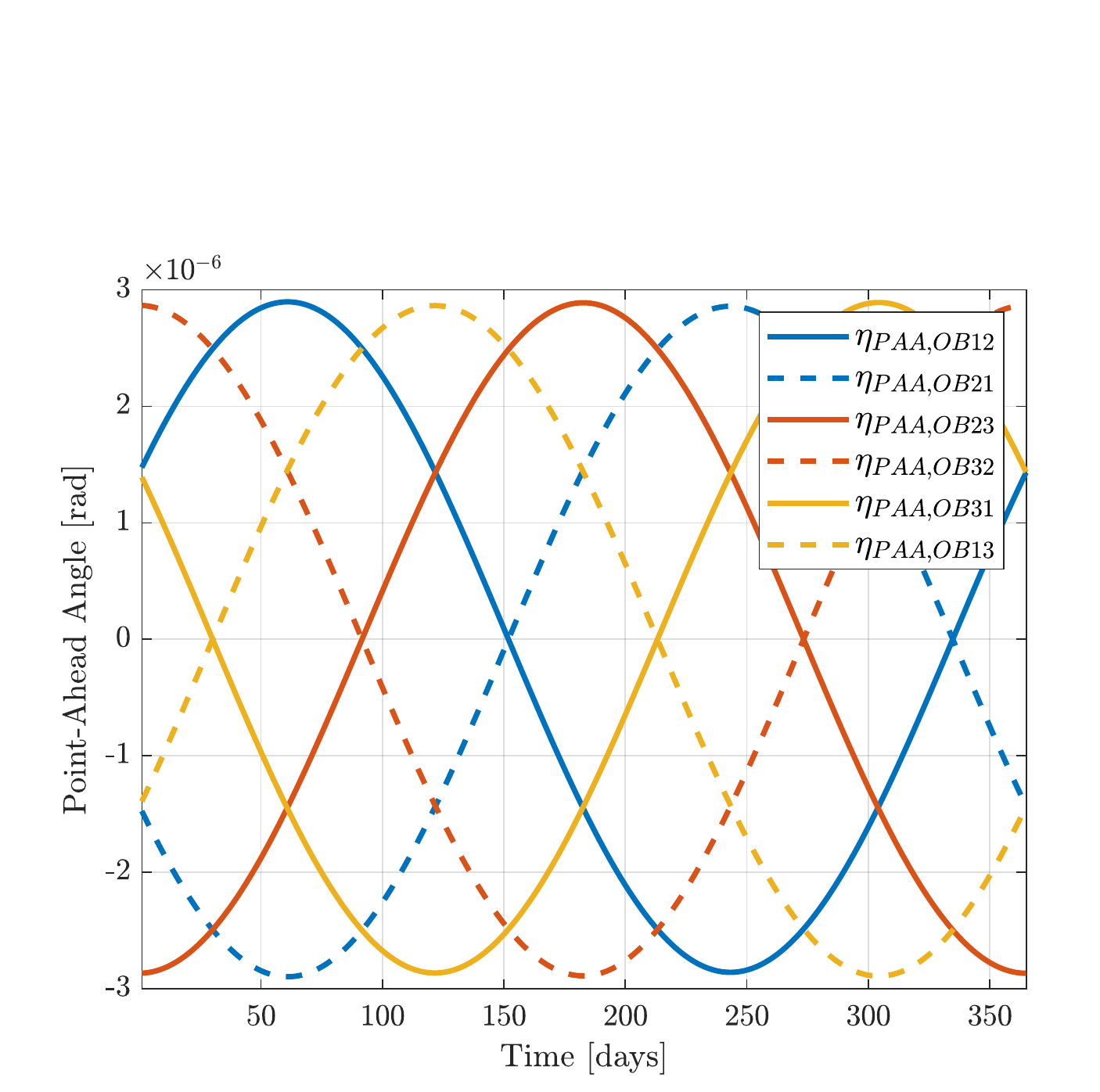}
  \caption{Out-of-Plane PAAs ($\kappa=\lambda=0$)}
  \label{im:OOPAA}
\end{minipage}%
\begin{minipage}{.5\textwidth}
  \centering
  \includegraphics[trim=90 260 100 180, clip,width=1\textwidth]{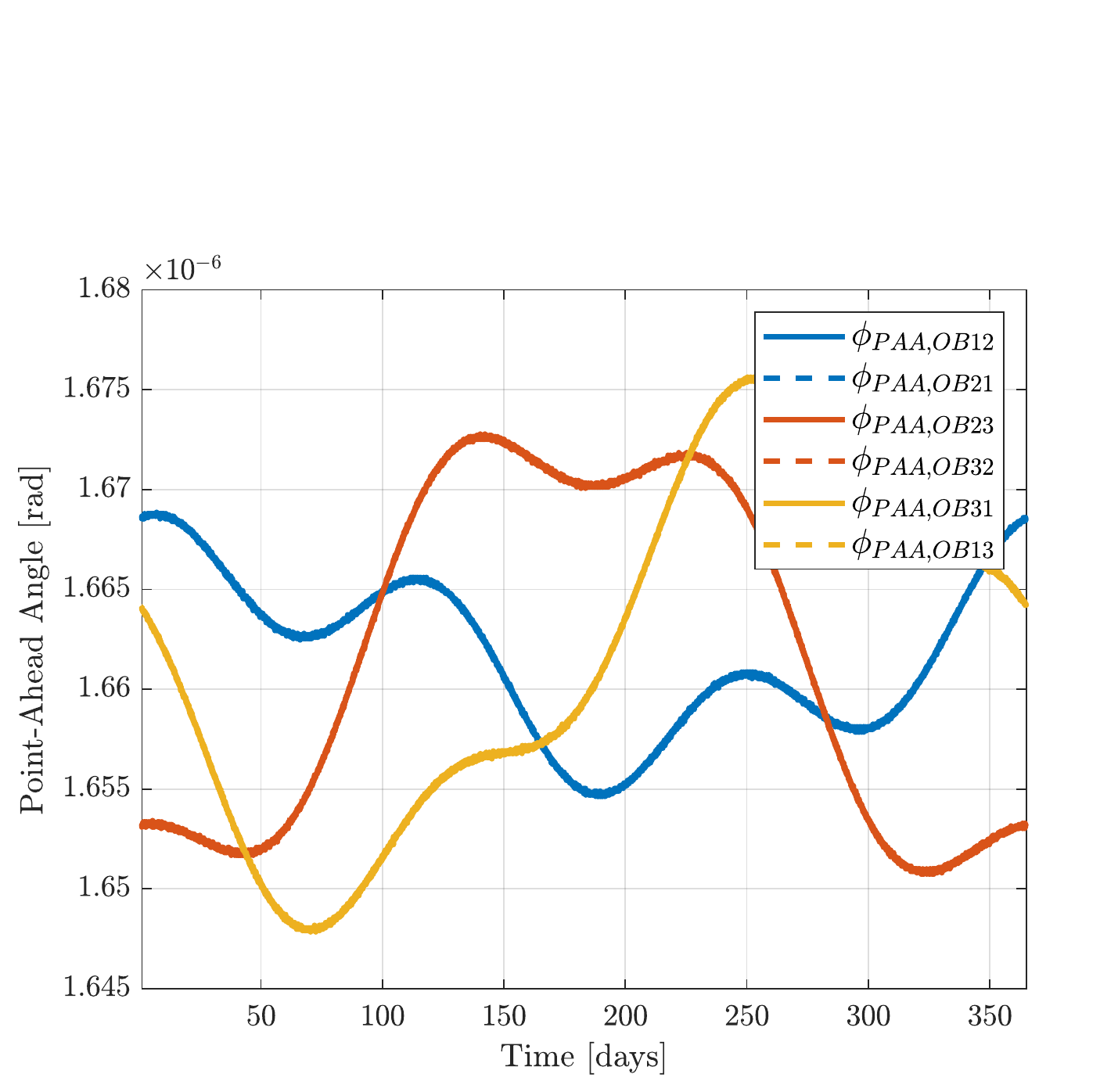}
  \caption{In-Plane PAAs ($\kappa=\lambda=0$)}
  \label{im:IPPAA}
\end{minipage}
\end{figure}

The evaluation of Eq. \eqref{eq:oopPAA} and Eq. \eqref{eq:ipPAA} based on Eqs. \eqref{eq:orbx} to \eqref{eq:orbz} is given in Figures \ref{im:OOPAA} and \ref{im:IPPAA} for an orbital period of one year and $\kappa=\lambda=0$. The impact of $\kappa,\lambda\neq 0$ on PAA control robustness is addressed at the end of the paper.  From Figure \ref{im:OOPAA}, the out-of-plane PAA has a peak-to-peak amplitude of $6.8 \,\mathrm{{\upmu} rad}$ oscillating around almost zero. Note that the out-of-plane PAA of opposite optical benches are identical with reversed sign, whereas the in-plane PAA of opposite optical benches match in sign.  From Figure \ref{im:IPPAA}, the PAA mean in-plane is $1.662\,\mathrm{{\upmu} rad}$ with an average peak-to-peak amplitude of only $0.0214\,\mathrm{{\upmu} rad}$.  It is only necessary to readjust the out-of-plane PAA, not the in-plane one. This is because in-plane PAA amplitude of $20\,\mathrm{nrad}$  is almost negligible compared to the half-angle of the beam ($\approx 2\,\mathrm{{\upmu} rad}$), and so is the resulting intensity loss.
\section{Optimal Point-Ahead Angle Control}\label{sec:3}
\vspace{5pt}
The section presents an open-loop control strategy for discrete correction of LISA's PAA and concludes with a robustness analysis of the control concept regarding varying initial spacecraft positions. Beforehand, the difference between continuous and discrete PAAM operation is summarized.
\subsection{Continuous and Discrete Point-Ahead Angle Adjustment Conditions}
Two different PAAM operation modes are possible.
The one-axial PAAM can be operated continuously to ensure that the out-of-plane PAA is always maintained at the correct value. Note that continuous operation implies the PAAM step frequency to be higher than the lower LISA measurement bandwidth limit of 0.1 mHz. The critical concern would be the generation of noise effects in-band due to the motion of the PAAM mirror. 
In discrete operation, also denoted step-and-stare, the PAAM is left in a static position, and the discrete actuation periodically corrects the PAA variation. This would require defining a maximum allowable PAA misalignment. Since the outgoing laser beam is held static, the advantage of discrete operation is to minimize the impacts from dynamic effects. One factor in determining the allowable off-pointing for discrete operation is the loss of received power due to the depointing of the transmitting beam on the remote spacecraft. Recent analyses show that the change in the local slope of the far-field wavefront will be more significant since it affects the magnitude of TTL coupling in the transmitting beam path. Note that the long-arm TTL factors are expected to be a few millimeters, taking into account the telescope magnification \cite{PhysRevApplied.14.014030}. The allowable off-pointing depends on the actual wavefront error and how much of the long-term stability budget is allocated to this effect. In this paper, a maximum off-pointing budget of $\Delta\eta_{PAA,max}=15$  nrad is selected, as an example, compensating only for the large variations of the out-of-plane PAA. Noise resulting from non-correcting the smaller in-plane PAA variations is currently being covered within LISA's performance budget. 
\subsection{Strategies for Discrete Point-Ahead Angle Control}
The section presents three different strategies for out-of-plane PAA adjustment: Spacecraft-local PAA adjustment, simultaneous PAA adjustment of all optical benches, and constellation-coordinated PAA adjustment of selected ones. Each strategy is evaluated from the perspective of TTL estimation.
\paragraph{Spacecraft-Local Point-Ahead Angle Adjustment}
A method for reducing the PAA adjustment-related TTL coupling factor changes is to minimize the overall number of adjustments. The number of PAA adjustments is minimized for the constellation when minimizing adjustments on each optical bench. This implies that PAAM actuation on an optical bench occurs only when the off-pointing budget is reached regarding that optical bench. The condition for PAA adjustment of each OB $ij$ can be stated independently from the other optical benches in this case. It is given by:
\begin{equation}
    \int_{t_{k,ij}}^{t_{k+1,ij}}\frac{\partial\eta_{PAA,ij}}{\partial t}dt=\Delta\eta_{PAA,max}\label{eq:Strat1}
\end{equation}
with $t_{k,ij}$ defined as the time step $k$ of the last PAA adjustment of OB $ij$ and $t_{k+1,ij}$ defining the time step $k+1$ of the next PAA adjustment of OB $ij$.
\begin{figure}
  \centering
  \begin{minipage}[b]{0.45\textwidth}
 \includegraphics[trim=100 260 90 270, clip,width=1.1\textwidth]{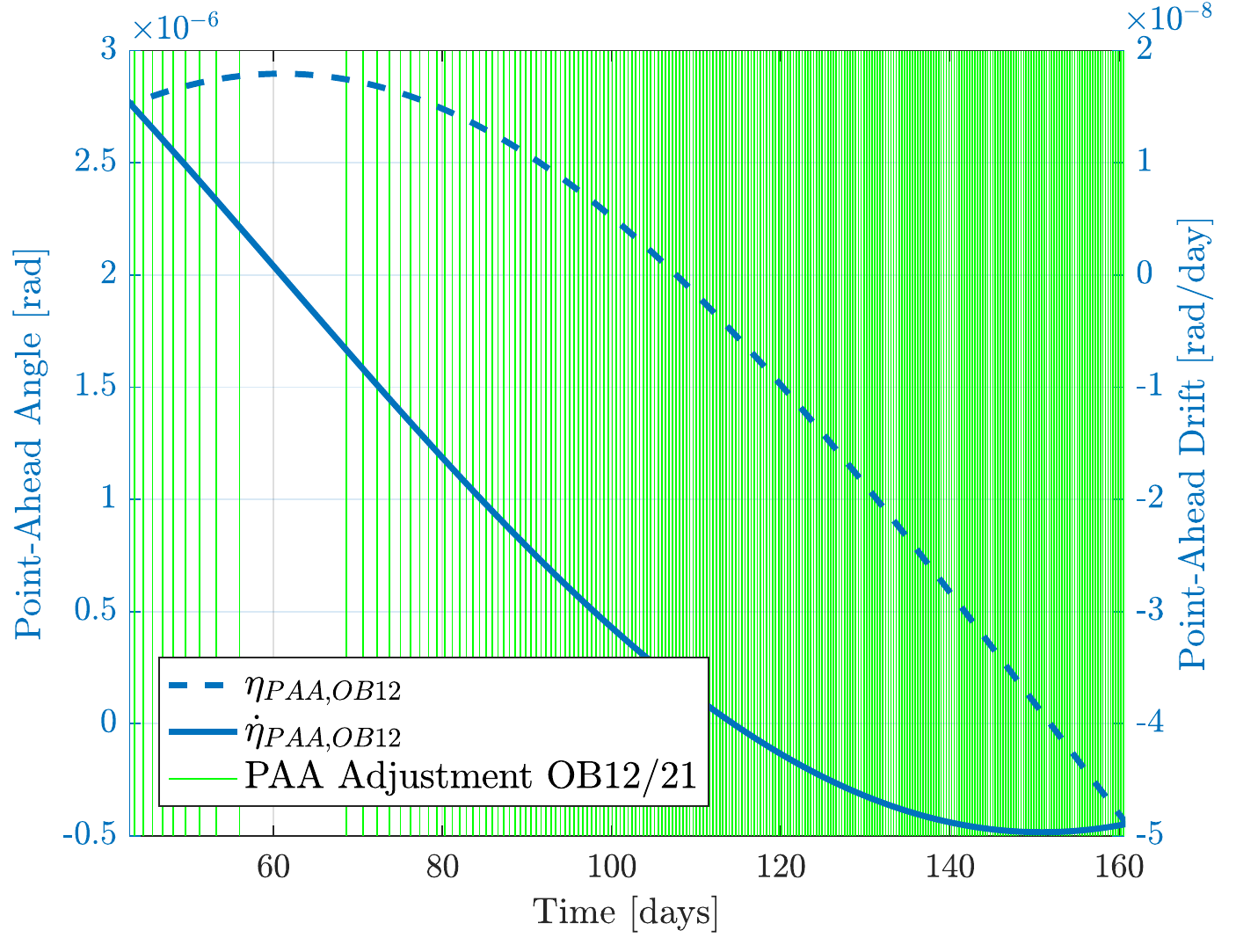}
    \caption{Discrete PAA12/21 Adjustments}\label{im:Strag1_ExOB12}
  \end{minipage}
  \hfill
  \begin{minipage}[b]{0.45\textwidth}
   \includegraphics[trim=90 260 100 270, clip,width=1.1\textwidth]{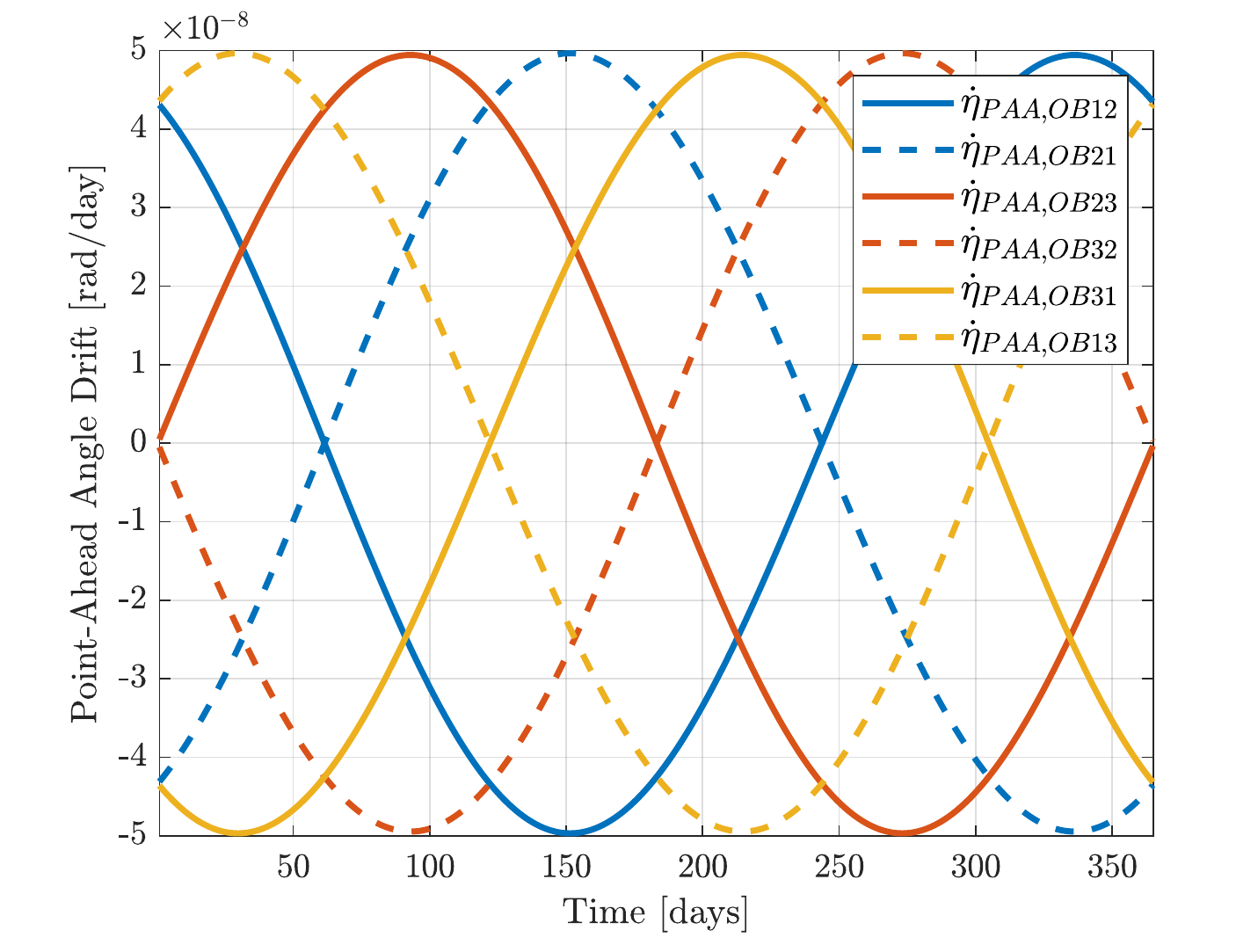}
    \caption{Out-of-Plane PAA Drifts}\label{im:OOPPAA_Drift}
  \end{minipage}
\end{figure}

Figure \ref{im:Strag1_ExOB12} presents the evaluation of Eq. \eqref{eq:Strat1} for 60 days using OB12 as an example. The green lines represent discrete PAAM actuations. Note that the out-of-plane PAA of OB12 is equal to the negative one of OB21. Therefore, the PAA adjustment condition is met simultaneously for the OB12/21 pair. Intervals between readjustments increase with decreasing PAA drift. Consequently, the maximum interval between two readjustments in Figure \ref{im:Strag1_ExOB12} is reached at the zero crossing of $\dot\eta_{PAA,OB12}$. It is about 12.65 days for the initial conditions $\kappa=\lambda=0$. The minimum time interval between readjustments corresponds to the maximum PAA drift. Figure \ref{im:OOPPAA_Drift} depicts the time derivatives of LISA's out-of-plane PAA. Considering a maximum rate of $50\,\textrm{nrad/day}$, the minimum time interval between subsequent adjustments is approximately 7.2 h.

\begin{figure}
  \centering
  \begin{minipage}[b]{0.45\textwidth}
 \includegraphics[trim=100 260 90 282, clip,width=1.1\textwidth]{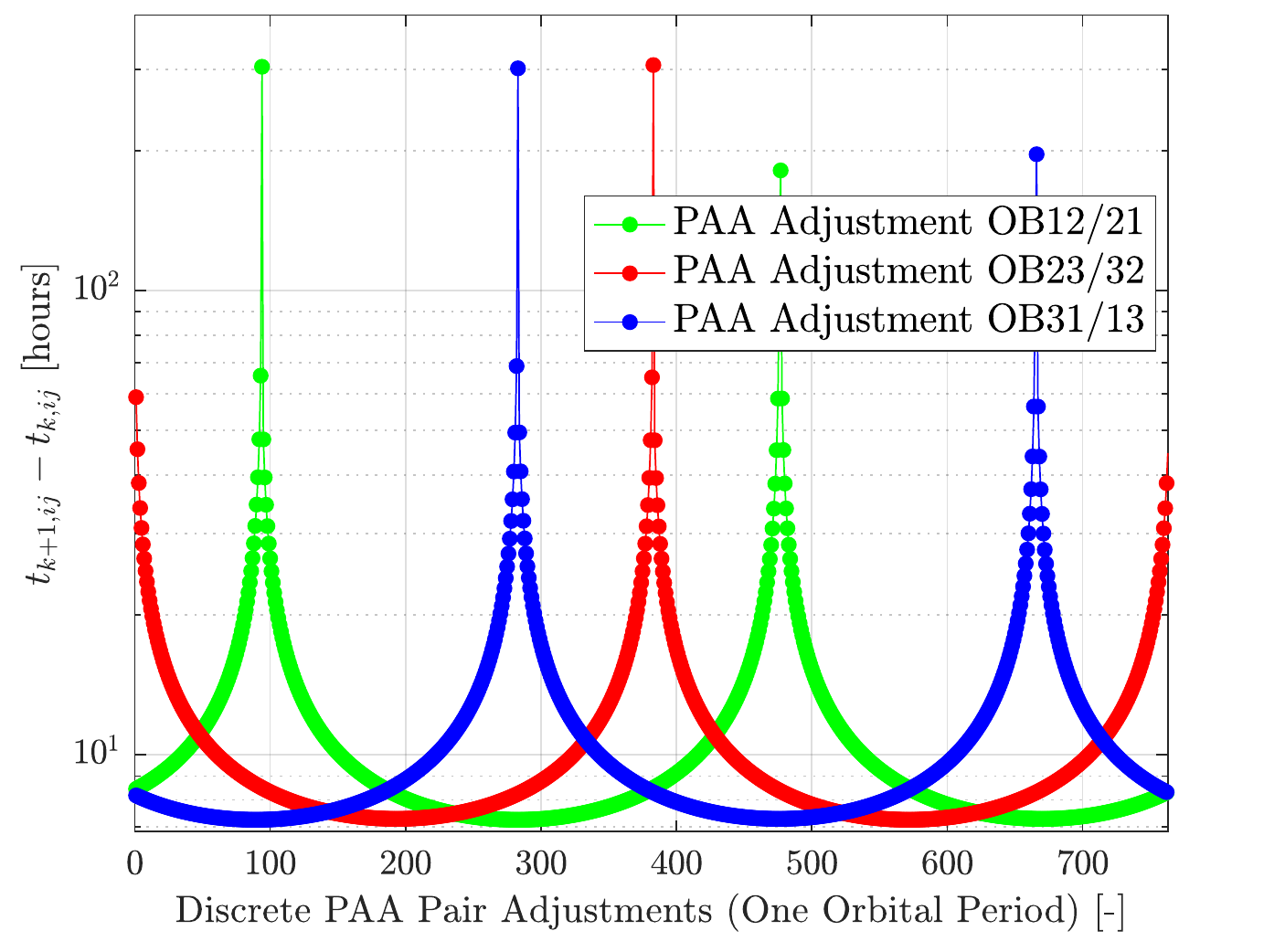}
    \caption{Spacecraft-Local PAA Adjustment}\label{im:Strat1_Overview}
  \end{minipage}
  \hfill
  \begin{minipage}[b]{0.45\textwidth}
   \includegraphics[trim=90 260 100 282, clip,width=1.1\textwidth]{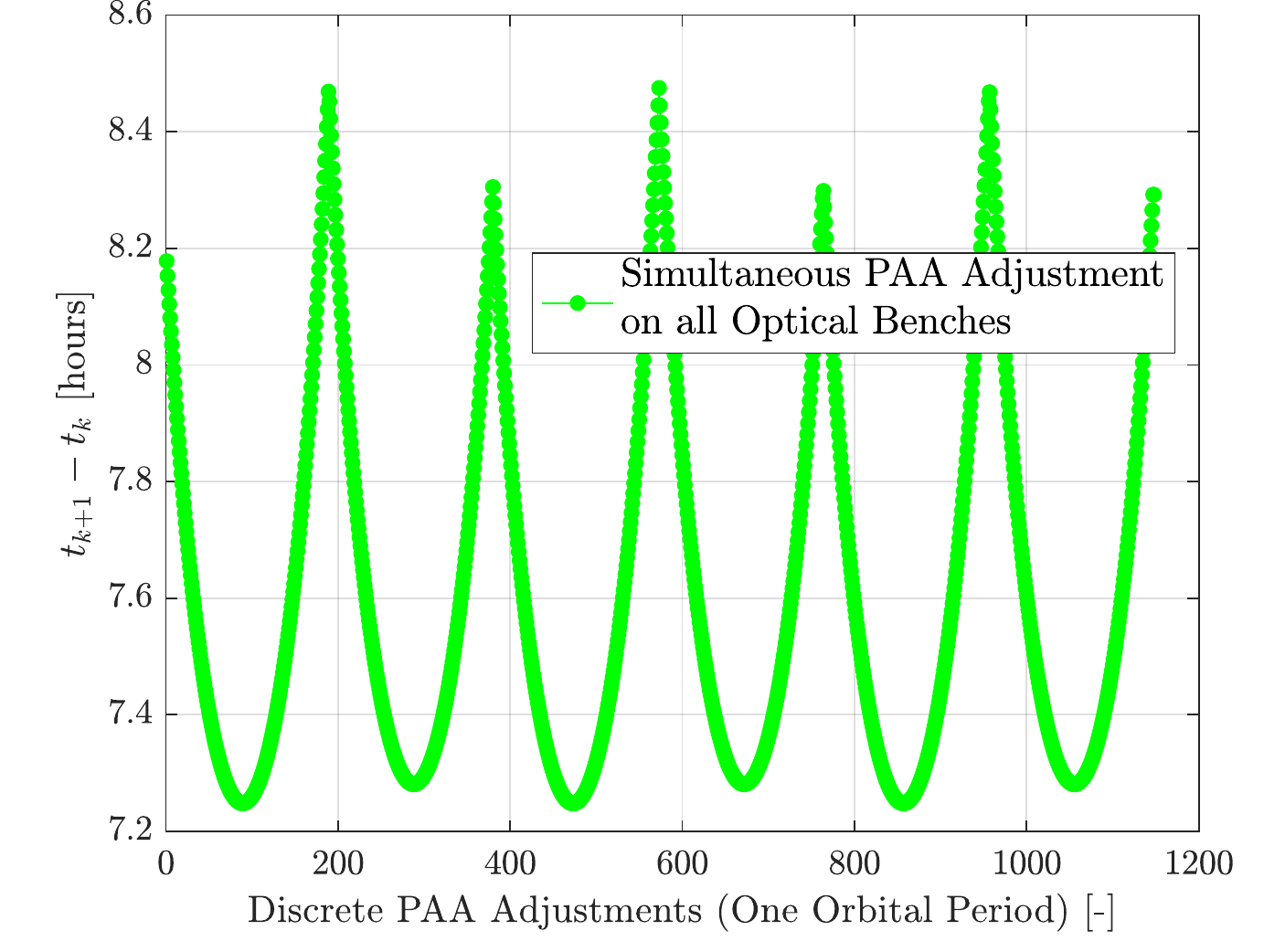}
    \caption{Simultaneous PAA Adjustment}\label{im:Strat2}
  \end{minipage}
\end{figure}

Figure \ref{im:Strat1_Overview} depicts time intervals between successive readjustments for all optical bench pairs and the entire LISA orbit. A total of 4590 readjustments are necessary during one orbital period for the selected initial conditions, representing the minimum number of PAAM actuation according to Eq. \eqref{eq:Strat1}. The drawback of the spacecraft-local PAA adjustment strategy is that the maximum and minimum readjustment intervals of the optical bench pairs are shifted to each other. Although the minimum PAA adjustment time is 7.2 h for one optical bench, the period between adjustments of different optical benches can be only a few seconds in the worst case. This is undesirable from the perspective of TTL noise estimation, since TTL noise estimation periods ideally do not contain a sudden jump of the TTL coupling factor induced by a PAA actuation \cite{houba}. With this in mind, minimizing the total number of PAA adjustments does not lead to a suitable operational concept for LISA.
\paragraph{Simultaneous Point-Ahead Angle Adjustment}
Time intervals between PAA adjustments of different optical benches can be increased, when all PAAM are activated simultaneously as soon as the maximum off-pointing budget is reached for one optical bench. The corresponding adjustment condition is:
\begin{align}
\begin{aligned}
t_{k+1}&= \min_{*} t_{k+1, i j}\label{eq:Strat2},\,\textrm{with}\\
* & \equiv \int_{t_{k}}^{t_{k+1, i j}} \frac{\partial \eta_{PAA,ij}}{\partial t} dt=\left.\Delta \eta_{P A A, \max }\right|_{i, j \in\{1,2,3\}, i \neq j}
\end{aligned}
\end{align} 
with $t_{k+1}$ defining the time step $k+1$ of the next PAA adjustment for all six optical benches.
Figure \ref{im:Strat2} presents the evaluation of Eq. \eqref{eq:Strat2}. Compared to the previous strategy, the time intervals between PAA readjustments no longer vary between a few seconds and several days. Instead, time intervals for simultaneous PAA adjustment lie between 7.2 h and 8.5 h generating more uniform TTL estimation windows. However, the number of readjustments increases by 50\% from 4590 to 6894 for $\kappa=\lambda=0$. Therefore, a combination of both strategies is presented in the following, indicating additional advantages for TTL estimation.
\paragraph{Constellation-Coordinated Point-Ahead Angle Adjustment} Periodic time intervals of up to 12 days were the main advantage of the spacecraft-local readjustment strategy. Unfortunately, time intervals of several days get lost in the simultaneous PAA adjustment approach. For this reason, an additional condition is added. Equation \eqref{eq:Strat2} still applies, defining the timestep $t_{k+1}$ at which the six optical benches can readjust their PAA with the difference of not having to. Equation \eqref{eq:Strat3} indicates whether $\eta_{PAA,ij}$ readjustment is required at $t_{k+1}$:
\begin{equation}
    \int_{t_{k,ij}}^{t_{k+2}}\frac{\partial \eta_{PAA,ij}}{\partial t} dt\geq \Delta\eta_{PAA,max}\label{eq:Strat3}
\end{equation}
The correction of PAA $ij$ is necessary at $t_{k+1}$ when the off-pointing between the last PAA readjustment at $t_{k,ij}$ and the next possibility at $t_{k+2}$ exceeds the allowed limit.

\begin{figure}[]
  \centering
 \includegraphics[trim=90 260 90 260, clip,width=0.5\textwidth]{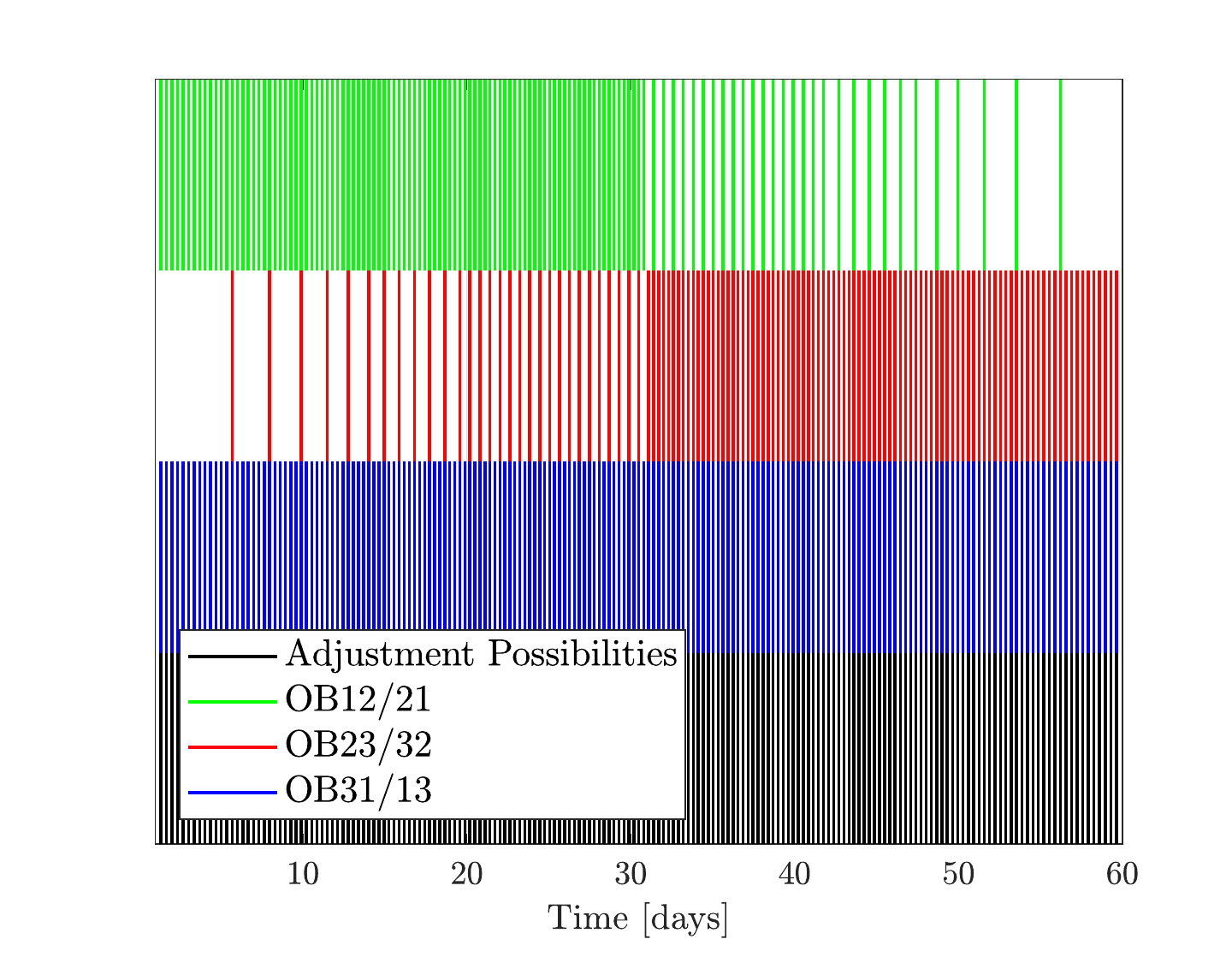} \caption{Constellation-Coordinated PAA Adjustment (Day 1 to Day 60)}\label{im:Strat3opseq60}
   \includegraphics[trim=90 260 100 260, clip,width=0.5\textwidth]{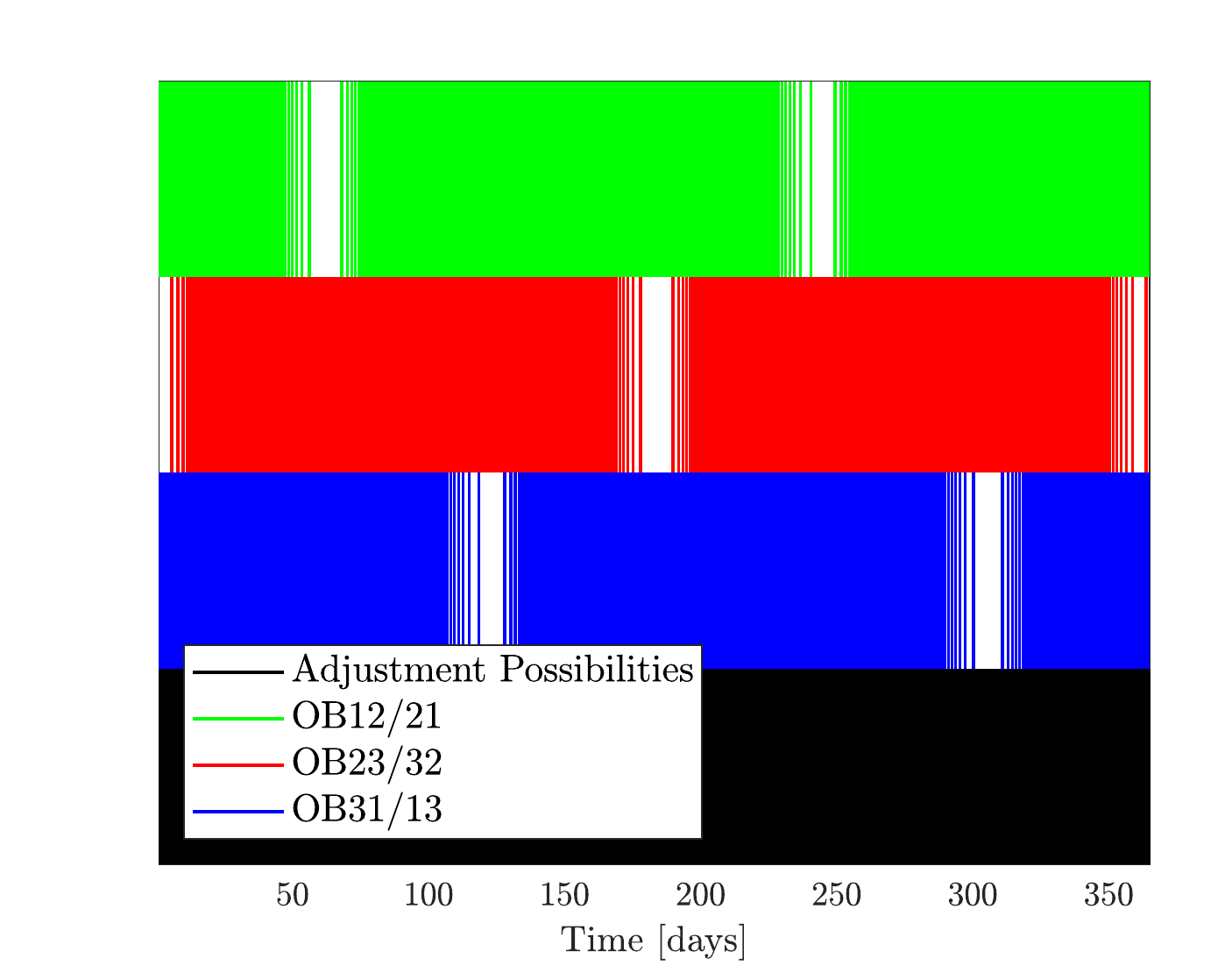}
   \caption{Constellation-Coordinated PAA Adjustment (Day 1 to Day 365)}\label{im:Strat3opseq265}
\end{figure}

Figures \ref{im:Strat3opseq60} and \ref{im:Strat3opseq265} show the operational sequence of PAA readjustments for the first 60 days and the entire LISA orbit of one year, respectively. According to Eq. \eqref{eq:Strat2}, black lines indicate timesteps when PAA corrections are possible. Colored lines denote actual adjustments of the optical bench pairs to be performed according to Eq. \eqref{eq:Strat3}. The total number of PAA readjustments required for $\kappa=\lambda = 0$ is 5284, 13\% more than for the local PAA readjustment strategy and 30\% less than for the simultaneous PAA readjustment strategy. The minimum time interval between consecutive PAA adjustments of different optical benches is 7.2 h, matching the results of the second strategy. Maximum time intervals between PAA adjustments of the same optical bench correspond to 12 days of the first strategy. The coordinated approach on the constellation level guarantees that only relevant PAA are adjusted. This allows for a periodic reduction of the TTL estimation equations because TTL coupling factors belonging to an optical bench that does not adjust its PAA in a specific interval are known from the previous estimation interval when neglecting short-term drifts.

\begin{figure}
  \centering
 \includegraphics[trim=90 260 90 260, clip,width=0.5\textwidth]{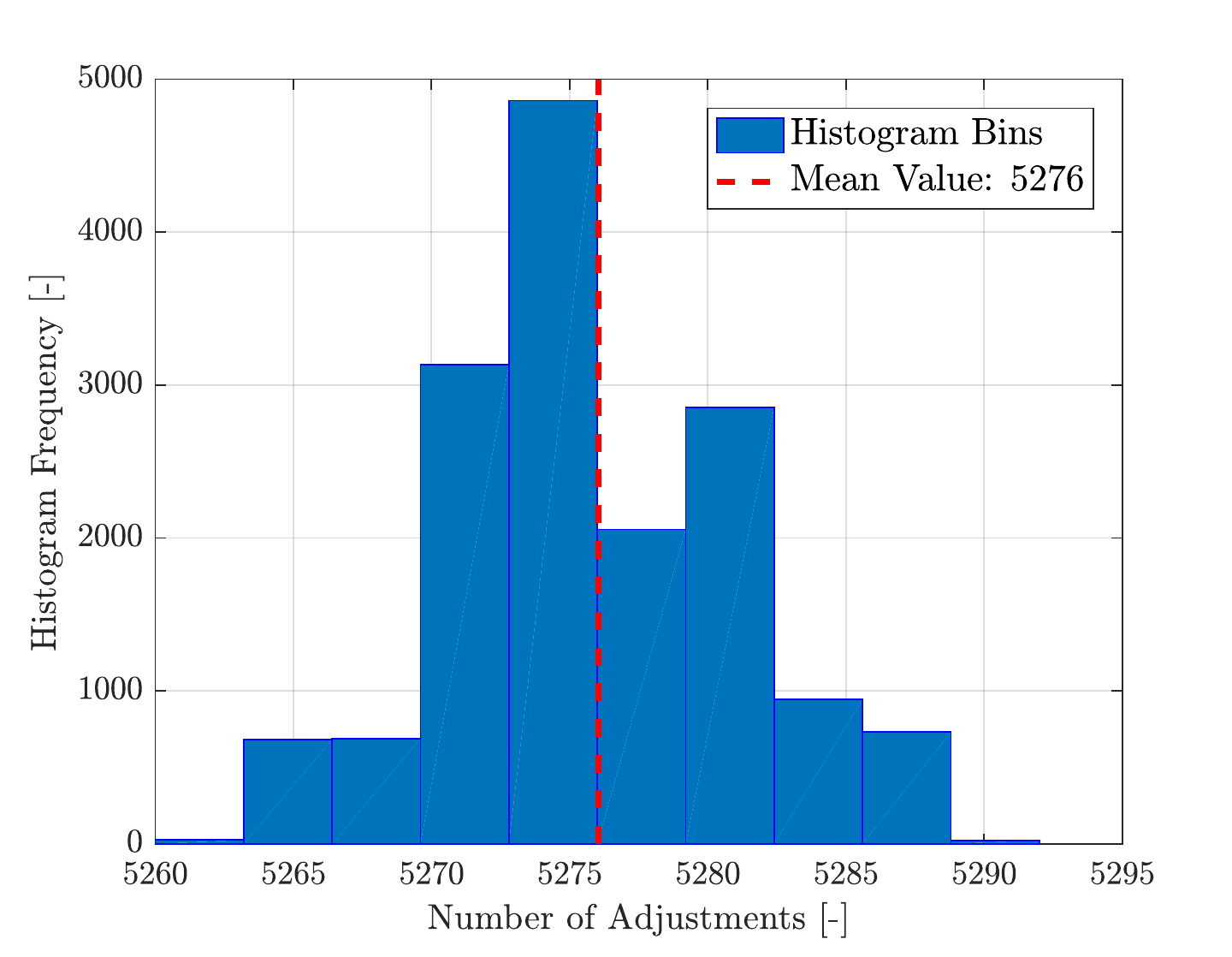} \caption{Total Number of PAA Adjustments for Different $\kappa,\lambda$}\label{im:Rob1}
   \includegraphics[trim=90 260 100 260, clip,width=0.5\textwidth]{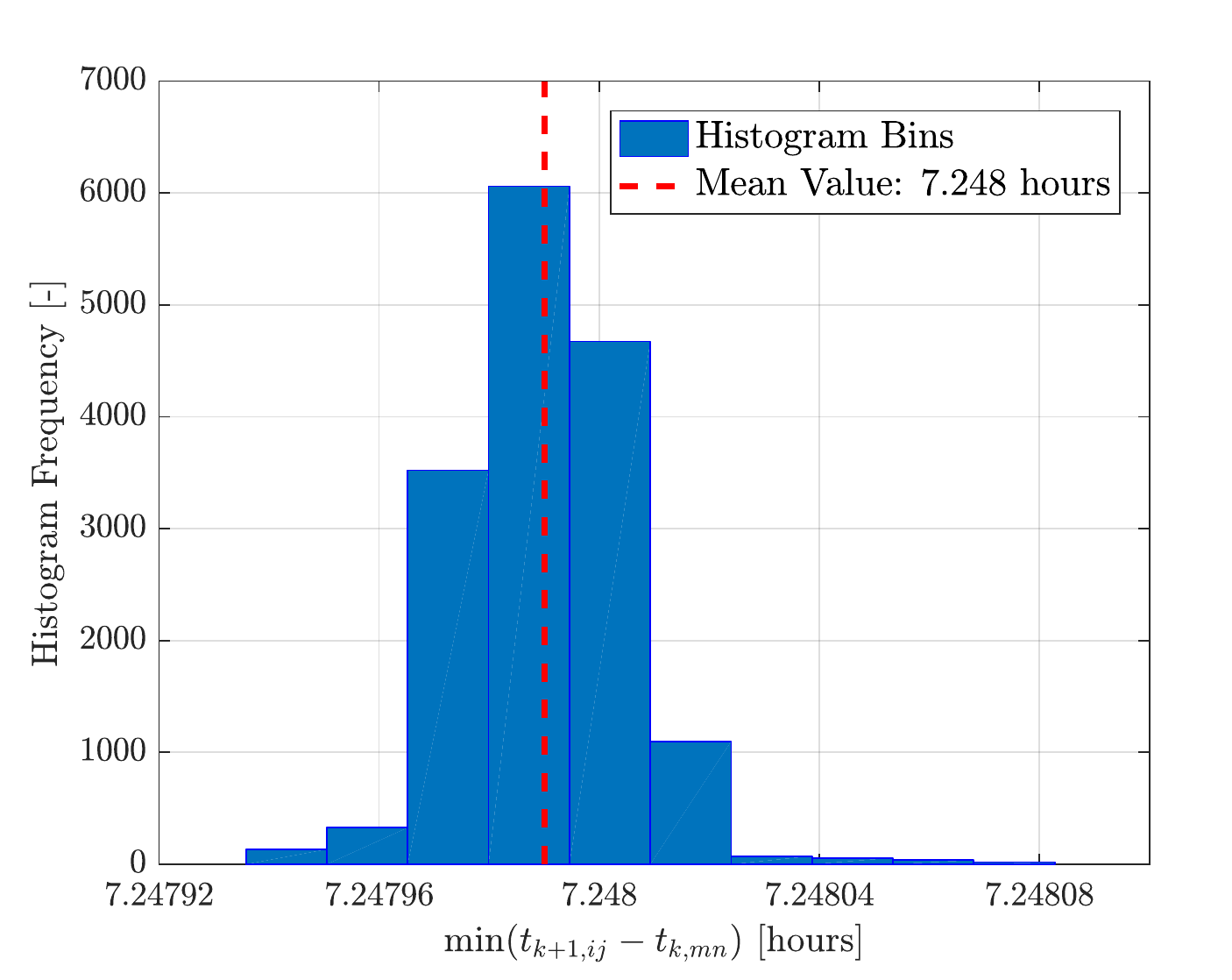}
    \caption{Minimum Time Interval between PAA Adjustments for Different $\kappa,\lambda$}\label{im:Rob2}
\end{figure}

The sensitivity of the total number of required PAA corrections regarding variations in parameters $\kappa$ and $\lambda$ is shown in Figure \ref{im:Rob1} for 15,000 different parameter combinations. The parameter $\kappa$ varies between $\pm\pi$, while $\lambda$ varies between $0$ and $2\pi$. The number of PAA readjustments ranges from 5260 to 5292 on constellation level and is insignificant. The same applies to the sensitivity of the minimum time interval between arbitrary optical benches OB $ij$ and OB $mn$ $\left(i,j,m,n\in 1,2,3\, \textrm{for}\, i\neq j,m\neq n\right)$, as shown in Figure \ref{im:Rob2}. The resulting time intervals differ by less than a second.  

A final overview of the three PAA open-loop control strategies is given in Table \ref{tab1:overviewStrat}.

\begin{table}
\centering
\begin{centering}
\caption{Overview of PAA Control Strategies $(\kappa=\lambda=0)$ }\label{tab1:overviewStrat} \vspace{10pt}
\begin{tabular}{cccc}
\hline
\hline
& \makecell{Spacecraft-Local \\ PAA Adjustment} & \makecell{Simultaneous \\ PAA Adjustment} & \makecell{Constellation-Coordinated \\ PAA Adjustment}\\
\hline
Adjustment Condition & Eq. \eqref{eq:Strat1} & Eq. \eqref{eq:Strat2} & Eqs. \eqref{eq:Strat2}, \eqref{eq:Strat3}    \\ 
\makecell{Number of Adjustments \\ per Year} & 4590 & 6894 & 5284  \\ 
\makecell{Minimum Time between \\ PAAM-related \\ TTL Factor Changes} &  \makecell{a few seconds \\ for a subset of factors} & \makecell{7.2 h \\ for all factors} & \makecell{7.2 h for all  \\ or a subset of factors}  \\
\makecell{Maximum Time between \\ PAAM-related \\ TTL Factor Changes} &  \makecell{12.65 days \\ for a subset of factors} & \makecell{8.5 h \\ for all factors} & \makecell{12.65 days \\ for a subset of factors}  \\
\hline
\hline
\end{tabular}
\end{centering}
\end{table}

\section{Conclusion} 
The paper presented an open-loop control strategy for point-ahead angle (PAA) correction, which is optimal from the perspective of estimating and subtracting tilt-to-length (TTL) coupling for the Laser Interferometer Space Antenna (LISA) mission. Therefore, analytical expressions were derived that describe temporal variations of the PAA, and a PAA adjustment condition is proposed guaranteeing that only relevant PAA are adjusted. The strategy allows for periodic reduction of unknowns in the mathematical formulation of the TTL estimator. Finally, the strategy's robustness towards variations in the initial orbit location and orientation of the LISA constellation was assessed. 



\section*{Acknowledgments}
The authors express their gratitude to Dr.-Ing. Georg Willich, the Head of Technical Strategy Space Germany at Airbus Defence and Space GmbH; Without his support, this research would not have been possible.

\bibliographystyle{AAS_publication}   
\bibliography{AAStemplatev2_0_6}   

\end{document}